\documentclass[dvips]{article}

\usepackage{icrc2011}

\title{Discovery of High-Energy and Very High-Energy Gamma-Ray Emission from the Blazar~RBS~0413}

\newcommand{\etal}{\MakeLowercase{\textit{et al. }}} 
\shorttitle{\c{S}ent\"{u}rk \etal HE and VHE Gamma-Ray Emission from RBS 0413}

\authors{G\"{u}ne\c{s} D. \c{S}ent\"{u}rk$^{1}$ for the VERITAS Collaboration$^{2}$, Pascal Fortin$^{3}$, Deirdre Horan$^{3}$ for the Fermi-LAT Collaboration}
\afiliations{$^1$Physics Department, Columbia University, New York, NY 10027, USA\\ $^2$see J. Holder et al. (these proceedings) or http://veritas.sao.arizona.edu/conferences/authors?icrc2011 \\ $^{3}$Laboratoire Leprince-Ringuet, Ecole Polytechnique, CNRS/IN2P3, Palaiseau, France}
\email{gunessenturk@gmail.com}

\abstract{We report on the discovery of high-energy (HE; $E>$ 0.1 GeV) and very high-energy (VHE; $E>$ 100 GeV) $\gamma$-ray emission from the blazar RBS~0413 with \emph{Fermi} Large Area Telescope (LAT) and VERITAS. The spectral energy distribution (SED), including contemporaneous X-ray and optical observations, is presented. Synchrotron self-Compton, external-Compton and lepto-hadronic models are applied to the SED and the results are discussed.}

\begin{document}
\maketitle

\section{Introduction}
Blazars are active galactic nuclei (AGN) that have their jet axes oriented at a small angle with respect to the observer. 
They are broadly classified as either flat-spectrum radio quasars (FSRQ) or BL Lacertae objects (BL Lac) according to their optical spectra, and are known to emit non-thermal radiation characterized by a double-peaked SED. 
The low-energy peak, covering the optical to UV/X-ray bands, is usually explained as due to  synchrotron emission from relativistic electrons in the blazar jets. 
The origin of the high-energy peak, occurring in the X-ray to $\gamma$-ray regime, is still not completely resolved and could be due to leptonic~\cite{ghisellini96} or hadronic~\cite{aharonian00} processes. 
RBS~0413 was discovered in the X-ray band (1E 0317.0+1834) during the Einstein Medium Sensitivity Survey and was optically identified as a BL Lac~\cite{gioia84}. 
The object was also detected as a radio emitter by the Very Large Array of the National Radio Astronomy Observatory~\cite{stocke90}. 
It has a redshift of 0.190~\cite{gioia84, stocke85}.
Having a ``featureless'' optical spectrum~\cite{stocke89} and an estimated synchrotron peak frequency $\mathrm{log}(\nu_\mathrm{peak}/1\mathrm{Hz})=16.99$~\cite{nieppola06}, RBS~0413 is classified as a high-frequency-peaked BL Lac object (HBL; $\nu_\mathrm{peak}>10^{15}\mathrm{Hz}$,~\cite{padovani96}).
The MAGIC Collaboration observed the object in 2004 - 2005 and reported a $\gamma$-ray flux upper limit of  $4.2\times10^{-12}~\textrm{ergs}~\textrm{cm}^{-2}~\textrm{s}^{-1}$, at 200\,GeV, assuming a $\Gamma=-3.0$ power-law spectrum~\cite{albert08}. 
VERITAS observed the source in 2008-2009 season and obtained a marginal significance of $\sim3$ standard deviations ($\sigma$). 
In 2009, \emph{Fermi}-LAT reported HE emission from the direction of RBS~0413~\cite{abdo2010}, triggering new VERITAS observations. 
In October~2009, VERITAS discovered $\gamma$-ray emission from RBS~0413~\cite{ong2009}. 

\section{Data Analysis and Results}

\subsection{VERITAS}

VERITAS is a ground-based VHE $\gamma$-ray observatory consisting of four atmospheric-Cherenkov telescopes, located at the Fred Lawrence Whipple Observatory (FLWO) near Amado, AZ, USA. It is sensitive to $\gamma$ rays with energies from 100 GeV up to 30 TeV, with an energy resolution of 15-25\%. 
In summer 2009, one of the telescopes was relocated and a new mirror-alignment system was introduced~\cite{mccann2010}, improving the sensitivity of the array~\cite{perkins09}. 
VERITAS can detect a $\gamma$-ray source with a flux of 1\% of the Crab Nebula flux in less than $\sim30$ hours, with angular resolution less than $0.1^{\circ}$ and pointing accuracy better than 50". 
Data analysis consists of a number of steps, including calibration, image parametrization, event reconstruction, background rejection and signal extraction, as described in~\cite{daniel08}.

Twenty-six hours of good-quality VERITAS data on RBS~0413 taken between Sep~2008 and Jan~2010 resulted in an excess of 108 candidate $\gamma$ rays and a 5.50\,$\sigma$ detection of the blazar. 
The spectrum above 250\,GeV can be described with a power law:
$dN/dE=F_0(E/E_0)^{-\alpha}$,
where $F_0=(1.38\pm0.52)\times10^{-7}$~$\textrm{TeV}^{-1}$~$\textrm{m}^{-2}$~$\textrm{s}^{-1}$, $\alpha=3.18\pm0.68_{stat}$ and $E_0 = 300$\,GeV, with a value of the $\chi^2$ per degree of freedom ($\chi^2$/dof) of 0.14/2, corresponding to a fit probability of $93\%$.
No evidence for variability is detected (see Figure\,1 top panel).

 \begin{figure}[!t]
  \centering
  \includegraphics[width=3.5in]{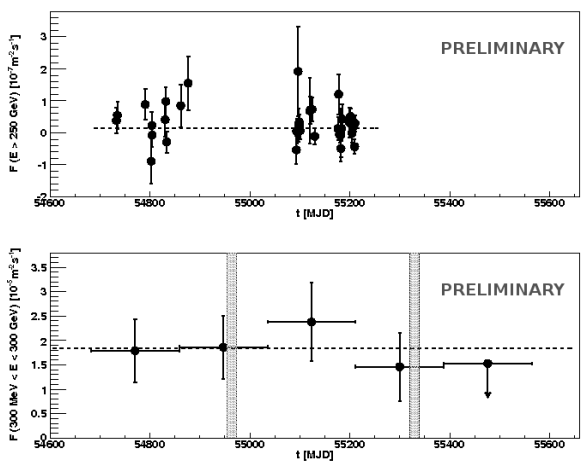}
  \caption{RBS 0413 VERITAS (\emph{top}) and \emph{Fermi}-LAT (\emph{bottom}) light curves. The dashed lines show the average flux. The shaded areas in the \emph{Fermi}-LAT light curve represent the time intervals excluded due to solar contamination.}
  \label{wide_fig}
 \end{figure}

\subsection{\emph{Fermi}-LAT}

\emph{Fermi}-LAT is a pair-conversion $\gamma$-ray detector sensitive to photons in the energy range from below 20 MeV to more than 300 GeV \cite{atwood2009}. 
The present analysis includes the data taken between 4~Aug~2008 and 4~Jan~2011, which covers the entire VERITAS observation interval. 
"\emph{Diffuse class}`` events with energy between 300\,MeV and 300\,GeV from a square region of side $20^{\circ}$ centered on RBS~0413 and with zenith angles $<100^{\circ}$ were selected for this analysis. 
The cut at 300\,MeV was used to minimize larger systematic errors at lower energies. 
The time intervals when the source was close to the Sun (MJD 54954-54974 and 55320-55339) were excluded.
The data were analyzed with the LAT Science Tools version v9r20p0\footnote{http://fermi.gsfc.nasa.gov/ssc/data/analysis/scitools/overview.html} and the post-launch instrument response functions P6\_V11\_DIFFUSE. 
The binned maximum likelihood tools were used for significance and flux calculation~\cite{cash79,mattox96}. 
Sources from the 1FGL catalog \cite{abdo2010} located within a square region of side $24^{\circ}$ centered on RBS 0413 were included in the model of the region. 
The background model includes the standard galactic and isotropic diffuse emission components\footnote{http://fermi.gsfc.nasa.gov/ssc/data/access/lat/BackgroundModels.html}.

A point source positionally consistent with RBS~0413 is detected with a significance of more than 9\,$\sigma$ (test statistic, $\mathrm{TS}=89$). 
The photon energy spectrum is  well described by a power law. 
Replacing the power-law model with a log-parabola model does not significantly improve the likelihood fit (the likelihood ratio test rejects the power-law model with a significance of only 0.35\,$\sigma$). 
The time-averaged integral flux is \emph{I} (300 MeV $< E <$ 300 GeV) = ($1.64 \pm 0.43_{stat}+0.31_{sys}-0.22_{sys}$) $\times10^{-9}$~$\textrm{cm}^{-2}$~$\textrm{s}^{-1}$. The spectral index is $1.57 \pm 0.12_{stat}$$+0.11_{sys}-0.12_{sys}$. 
The spectral points were calculated using the procedure presented in \cite{abdo2010}. 
In the energy range 100--300\,GeV, no detection was obtained ($\mathrm{TS}<9$) and an upper limit at the 95\% confidence level was derived. 
No evidence for variability was found when the data were analyzed using $\sim6$-month wide time bins (see Figure\,2 bottom panel).

\subsection{\emph{Swift}}

The VERITAS detection triggered a \emph{Swift}~\cite{Gehrels04} target-of-opportunity observation of RBS~0413 on 11~Nov~2009, with a total exposure of 2.4 ks. 
The \emph{Swift}-XRT photon spectrum is described by an absorbed power-law model with a fixed galactic hydrogen column density $N_{\rm{H}} = 8.91 \times 10^{20}$ cm$^{-2}$~\cite{Kalberla05}, yielding a $\chi^{2}$/dof value of 25.9/26. 
Over the energy range 0.3--10 keV, the best-fit photon index is $\Gamma = 2.22 \pm 0.07$, and the 1 keV normalization is ($3.31 \pm 0.22$) $\times 10$ keV$^{-1}$ m$^{-2}$ s$^{-1}$. No flux variability is evident over the 2.4 ks exposure.

UVOT observations were taken in the photometric band \emph{UVM2} (2246\,\AA) \cite{Poole08}. 
The \emph{uvotsource} tool was used to extract counts, correct for coincidence losses, apply background subtraction, and calculate the source flux. 
The standard 5" radius source aperture was used, with a 20" background region. 
The source flux was dereddened using the interstellar extinction curve in \cite{Roming09}. 
The measured flux is ($2.75 \pm 0.11$) $\times 10^{-12}$ $\textrm{erg}$~$\textrm{cm}^{-2}$~$\textrm{s}^{-1}$.

\subsection{MDM}

The R-band optical data were taken with the 1.3\,m McGraw-Hill telescope at the MDM observatory on Kitt Peak, Arizona, between 10 and 13~Dec~2009. 
All frames were bias-corrected and flat-fielded using standard routines in IRAF~\cite{Barnes93}, and instrumental magnitudes of RBS~0413 and six comparison stars in the same field of view were extracted using DAOPHOT~\cite{massey92} within IRAF. 
Physical magnitudes were computed using the physical R-band magnitudes of the six comparison stars from the NOMAD catalog~\cite{nomad}, assuming that the magnitudes quoted in that catalog are exact. 
The data show up to $\sim30$\% daily flux variability, with an average magnitude of 17.27.
The magnitudes were then corrected for galactic extinction using extinction coefficients calculated following~\cite{schlegel98}, taken from NED\footnote{http://ned.ipac.caltech.edu/} and then converted into $\nu\mathrm{F}_\mathrm{\nu}$ fluxes.

 \begin{figure*}[th]
  \centering
  \includegraphics[width=5in,height=4in]{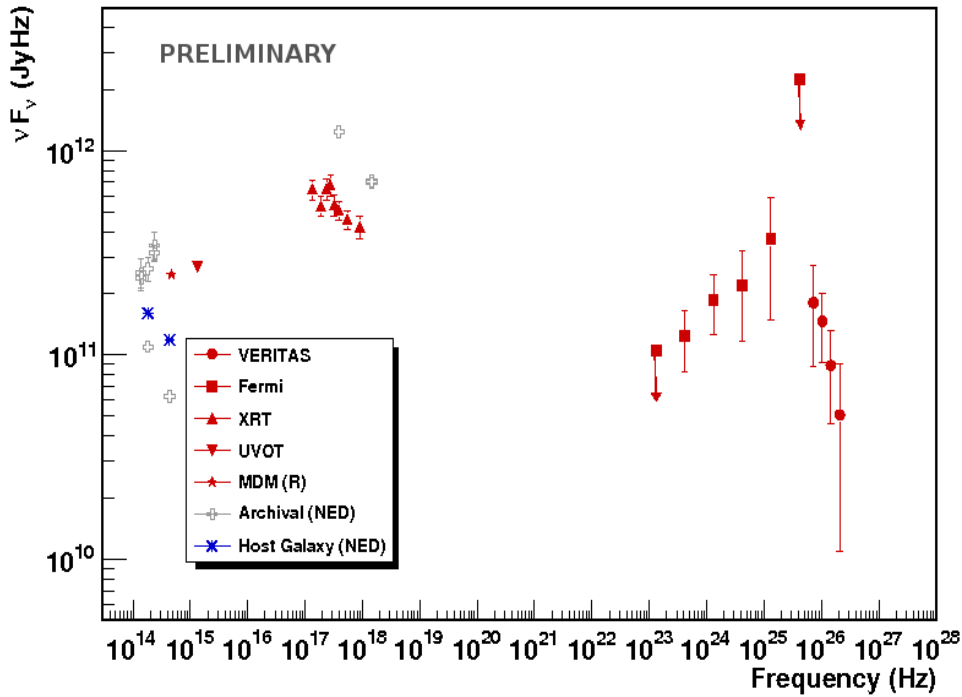}
  \caption{RBS 0413 spectral energy distribution.}
  \label{wide_fig}
 \end{figure*}

\section{SED Modeling}
The non-thermal continuum of RBS~0413 exhibits a double-peaked shape, typical for blazars.
In this study, three different time-independent models were used to describe the observed SED, using contemporaneous X-ray, UV and optical (R-band) data complementing the \emph{Fermi}-LAT and VERITAS observations (see Figure\,2). 
All model flux calculations were corrected for extragalactic background light (EBL) absorption using the model of~\cite{finke10}.
The first modeling, carried out assuming a pure SSC process, requires a magnetic field energy density which is only 6\% of that required for equipartition with the relativistic electron distribution.
In addition, the optical (R-band) spectrum is not reproduced, the model \emph{Fermi} spectrum is too hard and the model VERITAS spectrum is too soft, albeit within the errors in both cases.
Next, a combined SSC + EC model was adopted.
The addition of an EC component gives a better description of the optical and GeV parts of the SED with parameters very close to equipartition between the magnetic field and relativistic electrons, but slightly underproduces the TeV data.
Finally, a combined lepto-hadronic jet model as described in \cite{boettcher2010} was used. 
This model gives the best description for the data, with the system close to equipartition. 
On the other hand, it is the least constraining model given the large number of free parameters.

\section{Summary}
VERITAS and \emph{Fermi}-LAT detected VHE and HE $\gamma$-ray emission from the HBL RBS\,0413. 
No evidence for variability was seen in either data set.
The contemporaneous SED of RBS~0413 covering the energy range from the optical band to TeV $\gamma$ rays was studied and three different emission models were applied to the data. 
A pure SSC model fails to reproduce the R-band data and requires a magnetic field far from equipartition. The addition of an EC component yields an improved result but slightly underproduces the TeV data. 
The lepto-hadronic model gives the best description of the data, with the system close to equipartition, but it is the least constraining model since it has the largest number of free parameters.

The VERITAS research is supported by grants from the US Department of Energy, the US National Science Foundation, and the
Smithsonian Institution, by NSERC in Canada, by Science Foundation Ireland (SFI 10/RFP/AST2748), and by STFC in the UK. We acknowledge the
excellent work of the technical support staff at the FLWO and at the collaborating institutions in the construction and
operation of the instrument.

The \emph{Fermi}-LAT Collaboration acknowledges generous ongoing support from a number of agencies and institutes that have supported both the
development and the operation of the LAT as well as scientific data analysis.
These include the National Aeronautics and Space Administration and the Department of Energy in the United States, the Commissariat \`a l'Energie Atomique and the Centre National de la Recherche Scientifique / Institut National de Physique
Nucl\'eaire et de Physique des Particules in France, the Agenzia Spaziale Italiana
and the Istituto Nazionale di Fisica Nucleare in Italy, the Ministry of Education,
Culture, Sports, Science and Technology (MEXT), High Energy Accelerator Research
Organization (KEK) and Japan Aerospace Exploration Agency (JAXA) in Japan, and
the K.~A.~Wallenberg Foundation, the Swedish Research Council and the
Swedish National Space Board in Sweden.

Additional support for science analysis during the operations phase is gratefully
acknowledged from the Istituto Nazionale di Astrofisica in Italy and the Centre National d'\'Etudes Spatiales in France.


\clearpage

\end{document}